\documentclass[prd,amsmath,amssymb]{revtex4}
\newcommand{\met}{\rm\,/\!\!\!\!E_{T}}

\usepackage{graphicx}% Include figure files
\begin{document}

\begin{flushright} 
CDF/PUB/EXOTIC/PUBLIC/7265 \\
FERMILAB-CONF-04-307-E \\
version 2.0 \\
\today
\end{flushright}

\title{Neutral Higgs Boson Search at Tevatron \\ \vspace{1.0 cm} }

\author{Wei-Ming Yao\\
for the CDF and D0 Collaboration}
\affiliation{\vspace{0.5 cm} \\Lawrence Berkeley National Laboratory, One Cyclotron Road, Berkeley, CA 94720, 
USA\\E-mail: wmyao@lbl.gov}

\begin{abstract}
\vspace{1.0 cm} 

We review searches for neutral Higgs Boson performed by the CDF and D0 
collaborations using approximately 200 pb$^{-1}$ of the dataset accumulated from 
$p\bar p$ collisions at the center-of-mass energy of 1.96 TeV. No signals are found 
and limits on the Standard Model (SM) Higgs or SM-like Higgs production cross section times 
branching ratio and couplings of the Higgs boson in MSSM are presented, including 
the future prospects of discovery Higgs at the end of Run II.
\end{abstract}

\maketitle

\section{Introduction}
The great success of the Standard Model in explaining and predicting the experimental data provides
strong motivation for the existence of a neutral Higgs boson~\cite{sm}.  
With the recently improved top mass measurement~\cite{d0mass}, the global fit of 
electroweak precision data yields an estimation of the Higgs boson mass 
$m_h=113^{+60}_{-50}$ GeV/c$^2$ or $m_h \le 240$ GeV/c$^2$ at 95\% C.L.~\cite{ewg}, which agrees well with the 
direct searches from LEP 2 that set a limit on the Higgs mass $m_h\ge 114$ GeV/c$^2$ at 95\% C.L.~\cite{lep}.  
The Tevatron, the highest energy collider in the world, will have a window of opportunity to 
unlock the secrets of electroweak symmetry breaking (EWSB) before the LHC via either direct 
searches or precision measurements of 
the Top quark and W boson masses for better constraining the Higgs boson mass (indirect searches). 
In this note I will review the status of direct searches for neutral Higgs boson at the Tevatron. 

The upgraded CDF and D0 Run II detector are described elsewhere~\cite{det}. The Tevatron is doing very well and recently
has reached the record peak luminosity of $1\times 10^{32} cm^{-2}s^{-1}$, which is a design goal of Run II. 
Both CDF and D0 have recorded more than 450 pb$^{-1}$ of data up to August 2004 shutdown. The results presented 
here are mostly based on 200 $pb^{-1}$ up to Sept. 2003 shutdown.  

\section{Recent Run2 Results} 
Both CDF and D0 have re-established the top signal from Run II data, more importantly, the 
readiness of the tools required for lepton identification, $b$-tagging, jet clustering, and 
detector simulation. With these
tools and a well understood dataset in hand, many Higgs searches were performed. The results presented 
here are still at the engineering stage and much improved analyses with full datasets will emerge soon. 

\subsection{SM or SM-like Higgs Searches}
The experimental signature considered is  $ W h $ with $W \rightarrow
e\nu$ or $\mu\nu$, and $h\rightarrow b\bar b$,  giving final states with one
high-$P_T$ lepton, large missing transverse energy ($\met$)
due to the undetected neutrino, and two $b$ jets. 
The ability to tag $b$ jets using a secondary vertex detection 
with high efficiency and a low mistag rate is vital for searching for the
decay of $h\rightarrow b\bar b$. 
Both CDF and D0 select the $b$-tagged $W+$ 2 jet events since it is 
expected to contain most of the signal, while $b$-tagged $W+\ge 3$
jet events are  dominated by  $t\bar t$ decays.

\begin{figure}[htbp]
\begin{center}
\includegraphics[width=4.0in]{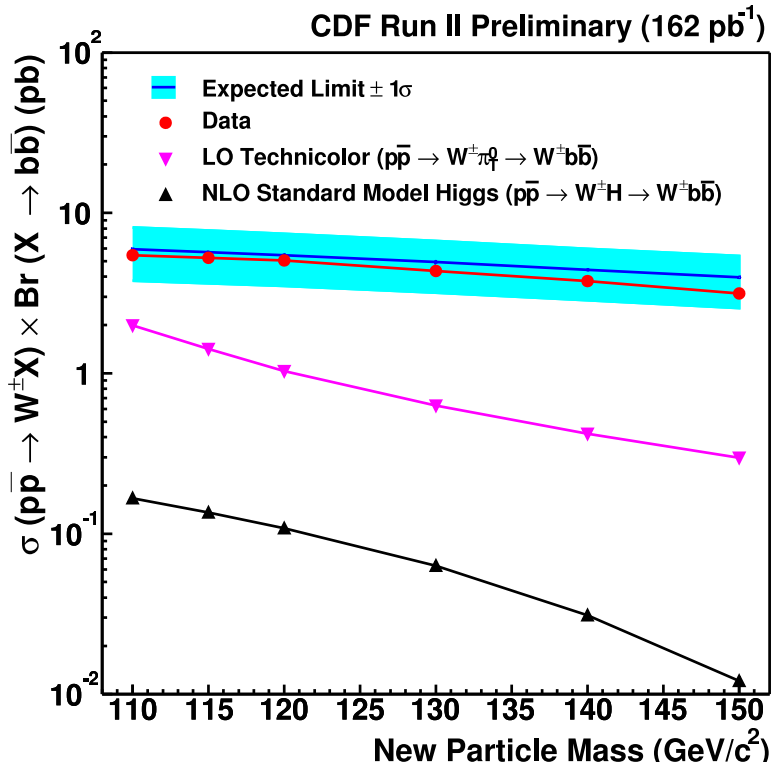}
\caption{The $95\%$ C.L. upper limit (red circle) on the $W^{\pm}H$ cross section
as a function of the Higgs boson mass. Also shown are the theoretical cross 
section (black triangle) for SM Higgs, the pseudo experiment results (blue line), and the
cross section of the Technicolor process (purple triangle).} \label{fig:bblimit}
\end{center}
\end{figure}

CDF observed 62 events with at least one $b$-tagged jet, consistent with 
the background expectation of $66\pm 9$ events, 
which are predominately from $Wb\bar b$, $Wc\bar c$, mistags, and $t\bar t$ decays. 
The likelihood fit to the mass distributions yields 
a limit at 95\% C.L. on the production cross section times branching ratio as a function of 
Higgs mass, 
shown in Figure~\ref{fig:bblimit}. The sensitivity of 
the present search is limited by statistics to a cross section approximately one  
order of magnitude higher than the predicted cross section for the Standard Model Higgs 
boson production, but is getting close to some of theoretical cross section for technicolor
particle production~\cite{tech}.

D0 also performed a similar search with double $b$-tagging using an integrated luminosity of 
approximately 174 pb$^{-1}$. 
They observed 2 events with an expectation of $2.5\pm 0.7$ and set a limit of 
12.4 pb at 95\% C.L. on the production cross section times branching ratio for the Higgs 
mass of 115 GeV/c$^2$.  They also searched for anomalous heavy-flavor decay in the lepton + jets 
sample containing both secondary vertex and soft lepton tags in the same jets (superjet). No 
significant deviation was found  with a cross section upper limit of 25.0 pb or 9.3 pb at 95\% C.L. 
for anomalous production of $Wb\bar b$-like or top-like events. 

\subsection{Search for $h\rightarrow W^+W^-\rightarrow l^+l^- \nu \bar \nu$} 
\begin{figure}[htbp]
\begin{center}
\includegraphics[width=4.0in]{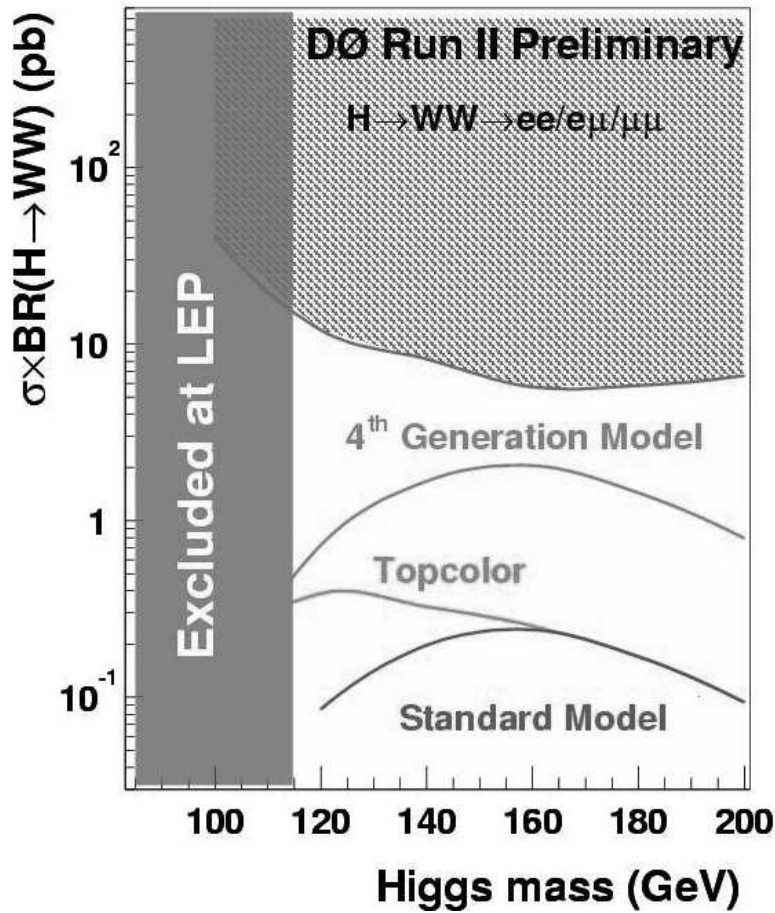}
\caption{The excluded cross section times branching ratio $\sigma \times BR(h\rightarrow W W^*)$ 
at 95\% C.L. together with expectations from Standard Model Higgs boson production and 
alternative models. } \label{fig:ww}
\end{center}
\end{figure}

For the Higgs mass above 130 GeV/c$^2$, the predominant decay mode of Higgs boson is to a pair of W bosons,
which offers an additional promising signature to look for Higgs by taking full advantage of large inclusive 
Higgs production with the decay $h\rightarrow W W^* \rightarrow l l \nu \nu $. We select the events with 
two opposite-sign high momentum lepton and large missing transverse energy. In order to reduce the 
significant $W W$ 
background, we exploit the spin correlations of $h\rightarrow W W^*$, which tends to produce the leptons 
close together. No signal is found and both CDF and D0 set a limit of 5.6 pb for the 
Higgs production cross section times branching ratio at 95\% C.L. as a function of Higgs mass, 
as shown in Figure~\ref{fig:ww}, along with predictions from the Standard Model and alternative 
models (Topcolor~\cite{topcolor} and 4th generation~\cite{4th}).  

\subsection{MSSM Higgs Searches} 

In the context of the minimal super-symmetric standard model (MSSM) the
Higgs sector has two doublets, one coupling to up-type quarks and the
other to down-type quarks.  There are five physical Higgs
boson states, denoted $h$, $A$, $H$, and $H^\pm$.  The masses and
couplings of the Higgses are determined by two parameters, usually
taken to be $m_A$ and $\tan\beta$ (the ratio of the vacuum expectation
value of the two Higgs doublets), with corrections from the scalar top
mixing parameters. In the case of large $\tan\beta$, 
there is an enhancement of $\tan^2\beta$ for
the production of $b\bar{b}\phi, \phi=h, A, H$ relative to the SM rate.  
This leads to distinct signature of four $b$ jets in the final states
including two b's from Higgs decay. 

\begin{figure}[htbp]
\begin{center}
\includegraphics[width=4.0in]{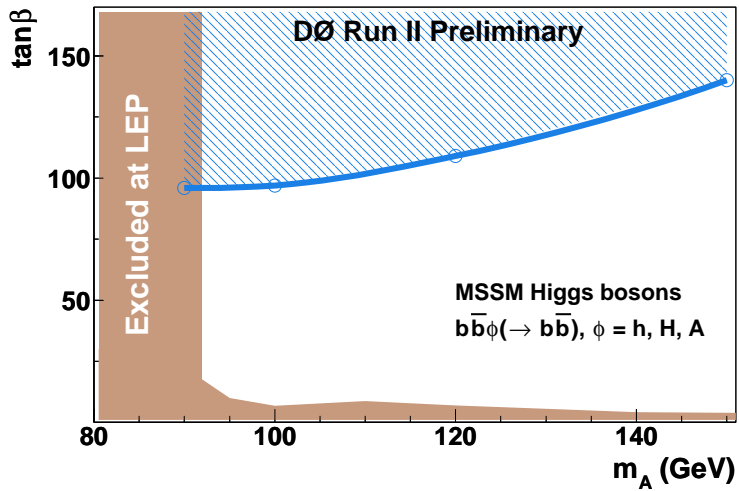}
\caption{The 95\% C.L. lower limit on $tan\beta$ set as a function of $m_A$ (thick, blue). The 
top hatched area indicates the excluded side of the line, and the shared area indicates the LEP
limits.}\label{fig:bbbb}
\end{center}
\end{figure}

D0 performed a search for neutral Higgs using approximately 130 pb$^{-1}$ of multi-jet sample.
After optimization, they select events containing 3 or 4 jets and required that 
at least three of jets be tagged. 
The invariant mass distribution of two leading $b$-tagged jets are consistent with the 
main backgrounds from QCD, fake tags, and $t\bar t$.
There is no evidence of signal found in the plot, which excludes the value of $tan\beta >80-120$ at 95\% C.L. 
in the region of $m_A$ between 90 and 150 GeV/c$^2$ in MSSM parameter space, as shown in Figure~\ref{fig:bbbb}. 

CDF also performed a search for a neutral MSSM Higgs decaying to tau pairs using a sample with 
an integrated luminosity of approximately 200 pb$^{-1}$, collected with the
dedicated Run II lepton + track triggers. 
The events are required to have one isolated lepton ($E_T>10 GeV$ for electron and $P_T>10$ GeV/c for muon) and 
one identified hadronic tau candidate. Lots of good work has gone into developing a more efficient  
tau finding algorithm, which has been cross checked using $W\rightarrow \tau \nu$ data. 
CDF finds no evidence of signal in the reconstructed visible di-tau 
mass distribution and is able to set an upper limits at 95\% C.L. on the product of Higgs production 
cross-section and its branching fraction to taus as a function of Higgs mass~\cite{amit}. 

\subsection{Other Higgs Searches: $h\rightarrow \gamma \gamma$} 
In the Standard Model the Higgs boson decays mostly to $b$-quark, W, or Z boson 
pairs depending on the mass range, 
while the branching fraction for $h\rightarrow \gamma\gamma$ is too small to 
be useful for probing SM Higgs at the 
Tevatron. However many extensions of the SM allow enhanced decay of $h\rightarrow \gamma \gamma$ largely due
to suppressed couplings of fermions, such as Fermiophobic Higgs~\cite{phobic} or 
Topcolor Higgs~\cite{topcolor} scenarios. 
The data used for this 
analysis were collected with the D0 detector, which corresponds to a total integrated 
luminosity of 191 pb$^{-1}$. 
The events are selected with two reconstructed EM objects with $E_T>25$ GeV in the Central Calorimeter (CC) or 
End Calorimeter (EC) in the detector $\eta$ range of $|\eta|<1.05$ and $1.5<|\eta|<2.4$, respectively. 
In addition,
the $P_T$ of the diphoton system is required to be above 35 GeV to reduce the di-jet background.  

The invariant mass of diphoton distribution for the data are consistent with the predicted backgrounds. 
In the absence of an evidence for a signal, D0 is able to set an upper 95\% C.L. limit on the diphoton 
branching ratio ($\approx 0.8$) as a function of Higgs mass for Fermiophobic and Topcolor Higgs models. 

\section{Future Prospects} 
In 2003, the CDF and D0 collaborations were asked by DOE to provide a new estimation of the Higgs 
Sensitivity based on current Run II detector performance~\cite{sensitivity}. 
The studies focus on a number of important 
improvements including the detectors, $b$-tagging, dijet mass resolution,  and the advanced analysis techniques. 

\begin{figure}[htbp]
\begin{center}
\includegraphics[width=4.0in]{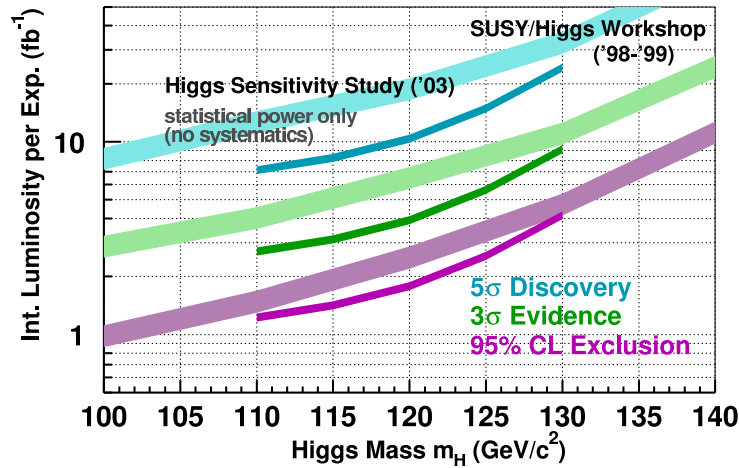}
\caption{The updated luminosity required for 95\% C.L. exclusion, 3 and 5 $\sigma$ discovery as 
 a function of Higgs mass. The effects of systematic uncertainties are not included yet.} \label{fig:lum}
\end{center}
\end{figure}

The updated integrated luminosity required to discover or exclude the SM Higgs, combining all search channels and
combining the data from both experiments, is shown in Figure~\ref{fig:lum}.  The finding is consistent with the 
SUSY-Higgs Workshop report, also shown in the plot~\cite{phobic}. We have not included the impact of systematic 
uncertainties in the curve yet. 
Controlling the systematic errors, especially dijet mass resolution will be important. 
The understanding of the Higgs sensitivity will improve over time once we get more data, a 
better understood detector, and more clever ideas, but 
finding Higgs at the Tevatron will be challenging. With 5 fb$^{-1}$ data, the Tevatron should be able to 
observe a 3 $\sigma$ 
excess for Higgs mass up to 120 GeV/c$^2$ or exclude the Higgs Mass up to about $m_h=130$ GeV/c$^2$ at 95\% CL 
if it is not present. The prospects for an MSSM Higgs is much better and 5 fb$^{-1}$ allows us to cover most of 
MSSM SUSY space for exclusion. 

\section{Conclusion} 

We have reviewed searches for neutral Higgs boson performed by the CDF and D0 
collaborations using approximately 200 pb$^{-1}$ of the dataset accumulated from 
$p\bar p$ collisions at the center-of-mass energy of 1.96 TeV. No signal is found 
and limits on the Standard Model (SM) Higgs or SM-like Higgs production cross section times 
branching ratio and couplings of the Higgs boson in MSSM are presented. 
In the next a few years, the Tevatron Collider is in an unique position to 
search for the dynamics responsible for electroweak symmetry breaking. 
We will be able to either see some glimmer of the new physics or constrain the 
Standard Model at an unprecedented level.

\section*{Acknowledgments} 
I would like to thank to the organizers of this excellent conference and their hospitality. 
It's my pleasure to thank all the people in the CDF and D0 collaborations whose work went into 
the results presented here.


\begin{thebibliography}{99}
    \bibitem{sm} 
    J. Gunion et al. ``The Higgs Hunter's Guide'' (Addison-Wesley; New York, 1990) 
    \bibitem{d0mass} 
    V.M. Abazov et al, Nature {\bf 429}, 638 (2004).     
    \bibitem{ewg} 
    The LEP EW Working Group (http://lepewwg.web.cern.ch/LEPWWG/) 
    \bibitem{lep} 
     A. Heister et al (The LEP Higgs Working Group), Phys. Lett. B {\bf 565}, 61 (2003). 
    \bibitem{det} 
    The CDFII Detector Technical Design Report, Fermilab-Pub-96/390-E
    The D0 Upgrade: The Detector and its Physics, Fermilab-Pub-96/357-E
    \bibitem{tech}
    K. Lane and S. Mrenna, Phys. Rev. D {\bf 67}, 115011 (2003). 
    \bibitem{topcolor} 
    C.T. Hill, PL B {\bf 266} 419 (1991); {\bf 345} 483 (1995). 
    \bibitem{4th} 
    E. Arik et al, EPJ C {\bf 26} 9 (2002). 
    \bibitem{amit} 
    A. Lath, New Physics Search at Tevatron, these proceedings. 
    \bibitem{phobic} 
    M. Carena et al (Higgs Working Group Report), hep-ph/0010338. 
    \bibitem{sensitivity}
    CDF and D$\rm{{\not}O}$ Collaborations, Results of the Tevatron Higgs Sensitivity Study,
    FERMILAB-PUB-03/320-E.
\end{thebibliography}
\end{document}